# Optimal Settings For Amplification And Estimation Of Small Effects In Postselected Ensembles

Aiham M. Rostom

aiham.rostom@gmail.com

Institute of Automation and Electrometry SB RAS, 630090, Novosibirsk, Russia.

**Abstract**

To describe the pre- and post-selected quantum ensembles, a complex quantity called the weak value of an operator is used. The weak value is highly controversial due to the fact that it is not bounded by the possible eigenvalues of the corresponding operator. Nevertheless, the obtaining of the anomalous weak value is regarded as a powerful technique in the quantum interferometry nowadays. Here it is shown that the postselection on a quantum system recovers a completely hidden interference effect in the measurement apparatus. Studying the interference pattern shows the optimal settings for the amplification and the parameter estimation. It also proves that the weak value is not an element of reality. Using single photons, it is investigated how a postselected photon can impart a $\pi$ phase shift (the peak of the amplification) to a photon interacting weakly with it in a nonlinear optical medium. The increasing of the degree of the entanglement lies behind the effectiveness of the postselection in the parameter estimation. In particular, arranging to postselect on pure entangled states can optimize the signal-to-noise ratio, allowing to achieve high-sensitive measurements using low input power.

## 1 Introduction

Pre- and post-selection of quantum states is an idea proposed by Aharonov, Bergmann and Lebowitz to reformulate the concept of quantum states in a time-symmetric manner [1]. It requires preparing the quantum system in identical initial (preselected) states and measuring it in particular final (postselected) states.

To describe pre- and post-selected ensembles when the intermediate interaction between the quantum system and the measuring device is weak, a complex quantity called the weak value of an operator was proposed [2].

The weak value can be anomalously large or small [2, 3, 4]. It is not bounded by the possible eigenvalues of the corresponding operator. For instance, the weak value of an operator $\hat{A}$ is defined by the formula: $\left\langle\hat{A}\right\rangle_w = \frac{\langle\psi^{post}|\hat{A}|\psi^{pre}\rangle}{\langle\psi^{post}|\psi^{pre}\rangle}$. By choosing almost orthogonal pre- and post-selected states $\left|\langle\psi^{post}|\psi^{pre}\rangle\right| \ll 1$, the weak value can take anomalously large values. This caused an interpretation controversy that continues up to the present time [5, 6, 7].

The weak-value amplification technique is the commonly used term in experiments when the obtaining of anomalously large weak values is the experimenter's "tactic" to amplify small effects of small parameters [4, 3, 8, 9]. The amplification manifests itself in the large shift of the measuring device when the postselection probability is low. In postselected quantum metrology, the amplification can be used to suppress "technical noise" and perform precise measurements [10]. For instance, a small spatial shift ($\sim A°$) resulting from the spin Hall effect of light [11], and angular mirror rotation of $\sim 400\, frad$ [12] have been measured using only milliwatts of input power.

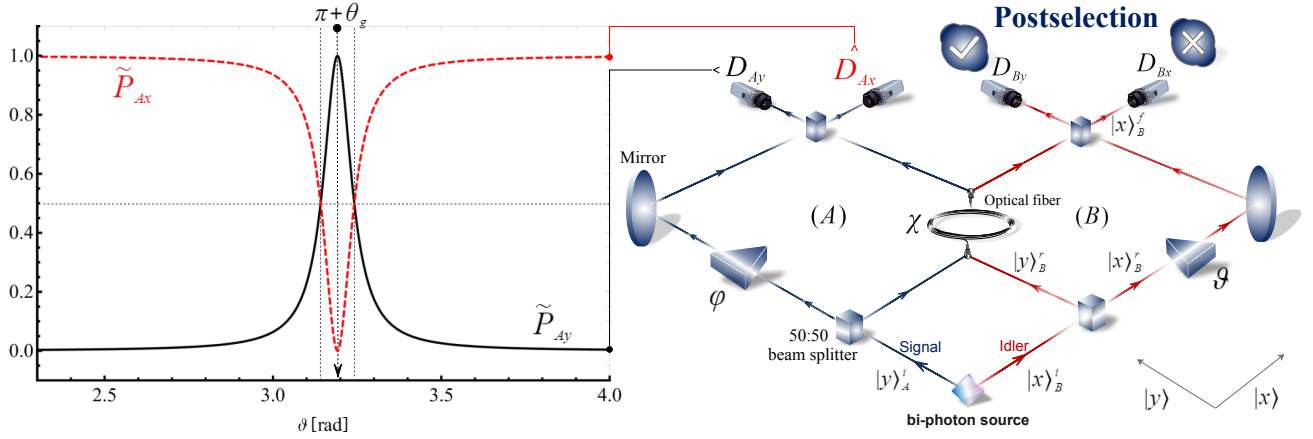

Figure 1: Illustration of the differential interferometry experiment. The right figure shows a system of two Mach-Zehnder interferometers coupled by cross-Kerr nonlinearity. B-interferometer is the system to be postselected on its output. A-interferometer represents the measuring device. The input state is a pair of time-energy correlated photons (signal and idler). By post-selecting one of the outputs of B-interferometer, upon the detection of idler photon in the detector $D_{By}$ for instance, the outputs of A-interferometer can be controlled by the reference phase $\vartheta$ applied in B-interferometer. For $\vartheta = \pi + \theta_g$, the postselection results from out-of-phase interference, and A-interferometer measures a switching effect between detectors $D_{Ax}, D_{Ay}$. The probabilities $\widetilde{P}_{Ax}, \widetilde{P}_{Ay}$, in the left figure are found from Formulae 9, for a relatively small value of the interaction strength of the cross-Kerr nonlinearity $\chi = \frac{1}{10} rad \ll \pi$.

Despite this effectiveness, some recent studies have also reached controversial conclusions about weak values and the role of discarding data (by postselection) in improving measurements [13, 14, 15, 16]. Other studies debate whether the weak-value amplification technique is a purely quantum mechanical concept, or it has a classical equivalent [17, 18].

The central message of this work is to introduce a new effect that can describe the quantum mechanics of pre- and post-selected ensembles without weak values, and to investigate the advantages it gives over the weak-value amplification in the interpretation, and hence, in the amplification issues and estimation of small parameters.

In sections 2 and 3, both standard and nonstandard interferometry (under postselection) are studied. A system consisting of two single-photon Mach-Zehnder interferometers coupled by a cross-Kerr nonlinearity is considered. The interferometers are denoted by A and B. Unlike the standard case, when the output of B-interferometer is postselected, the output interference of the A-interferometer (which represents the measuring device) can be controlled by a phase shifter applied on B-interferometer. This property is purely quantum mechanical, it contradicts all our classical intuitions about interference effects and can be understood as a new type of quantum erasers.

Next, in section 4, we show how a single postselected photon in B-interferometer can impart an exact $\pi$ phase shift (the peak of the amplification) to the other photon moving in A-interferometer, regardless of the cross-Kerr nonlinearity strength. The result is compared with previous experimental and theoretical works on the amplification of the single-photon nonlinearity using the weak-value amplification.

Section 5 deals with the parameter estimation issues. The imparted $\pi$ phase shift to the internal state of the A-interferometer can purify the entanglement between the output states of Mach-Zehnder interferometers. The pure entangled state lies behind the effectiveness of the quantum postselection-based metrology and serves as a resource for an optimal parameter estimation in the present paper. To validate, we describe how Fisher information gained by A-interferometer can be increased by an amount equal to the mean number of quanta entering B-interferometer. This optimizes the signal-to-noise ratio and guaranties the achieving of ultra-small uncertainty in the phase estimation using extremely low input power.

## 2 Standard interferometry

In the weak-value amplification technique, the postselection probability is determined by taking the square modulus of the weighting function in the denominator of the weak value $\langle \hat{A} \rangle_w = \frac{\langle \psi^{post}|\hat{A}|\psi^{pre}\rangle}{\langle \psi^{post}|\psi^{pre}\rangle}$, i.e., the postselection probability is $|\langle \psi^{post}|\psi^{pre}\rangle|^2$ [14]. However, this probability is only an approximation, since it contains no information about the interaction between the postselected quantum system and the measurement apparatus.

In this work, we use a different strategy to find the postselection probability. The exact postselection probabilities can be found only from the standard quantum mechanics (when all experimental runs are taken into account).

As an example, let us consider a pair of time-energy correlated photons (signal and idler) as preselected states of two Mach-Zehnder interferometers (MZI) $A$ and $B$. The photons traveling in $x-$ and $y-$directions are denoted by state vectors $|x\rangle$ and $|y\rangle$, see Figure 1. The balanced beam splitters of MZIs cause the following transformations

$$\hat{a}^i_{\sigma x} = \frac{1}{\sqrt{2}}\left(\hat{a}^r_{\sigma x} - \hat{a}^r_{\sigma y}\right), \hat{a}^i_{\sigma y} = \frac{1}{\sqrt{2}}\left(\hat{a}^r_{\sigma x} + \hat{a}^r_{\sigma y}\right), \tag{1}$$

$$\hat{a}^r_{\sigma x} = \frac{1}{\sqrt{2}}\left(\hat{a}^f_{\sigma x} - \hat{a}^f_{\sigma y}\right), \hat{a}^r_{\sigma y} = \frac{1}{\sqrt{2}}\left(\hat{a}^f_{\sigma x} + \hat{a}^f_{\sigma y}\right), \tag{2}$$

where $\sigma$ denotes to A or B interferometers, and the superscripts ($i, r$ and $f$) denote the initial, intermediate and final states respectively, $\hat{a}$ and $\hat{a}^\dagger$ are the annihilation and creation operators. The input state can be represented by two single-photon Fock states: $|y\rangle^i_A \otimes |x\rangle^i_B = \left(\hat{a}^i_{Ay}\right)^\dagger \left(\hat{a}^i_{Bx}\right)^\dagger |vac\rangle_A \otimes |vac\rangle_B$.

On the reference arms of A- and B-MZIs, controlled phases $\varphi$ and $\vartheta$ are applied. In the probe arms (in which a non-linear optical medium is placed), the signal and idler photons gain a cross-phase through Kerr effect in a nonlinear optical material.

The nonlinear interaction between photons is governed by the interaction Hamiltonian [19]: $\hat{\mathcal{H}} = \kappa\hbar\hat{n}^r_{Ax}\hat{n}^r_{By}$, where $\hat{n}^r_{Ax}$ and $\hat{n}^r_{By}$ are the photon number operators in the nonlinear medium in A- and B-MZIs respectively, and $\kappa$ is a constant proportional to the third-order susceptibility $\chi^{(3)}$ of the nonlinear material. The action of the medium can be represented by the unitary operator $\hat{U} = \exp\left[-\frac{i}{\hbar}t\hat{\mathcal{H}}\right]$. The unitary operator causes the following mode transformations in the intermediate states [19]

$$\hat{a}^r_{Ax} \to \hat{a}^r_{Ax}\exp\left[-i\chi\hat{n}^r_{By}\right], \hat{a}^r_{By} \to \hat{a}^r_{By}\exp\left[-i\chi\hat{n}^r_{Ax}\right], \tag{3}$$

where $\chi = \kappa t$ is the cross-phase, and $t$ is the interaction time.

Accordingly, the evolution of the preselected state $|y\rangle_A^i \otimes |x\rangle_B^i$, in MZIs gives the following output entangled state

$$|F\rangle_{AB} = \frac{1}{4}\left[\alpha |x\rangle_A^f |x\rangle_B^f + \beta |x\rangle_A^f |y\rangle_B^f + \gamma |y\rangle_A^f |x\rangle_B^f + \delta |y\rangle_A^f |y\rangle_B^f\right], \tag{4}$$

where $\alpha = e^{i\vartheta} - e^{i\chi} + e^{i(\vartheta+\varphi)} - e^{i\varphi}$, $\beta = e^{i\vartheta} + e^{i\chi} + e^{i(\vartheta+\varphi)} + e^{i\varphi}$, $\gamma = e^{i\vartheta} - e^{i\chi} - e^{i(\vartheta+\varphi)} + e^{i\varphi}$, $\delta = e^{i\vartheta} + e^{i\chi} - e^{i(\vartheta+\varphi)} - e^{i\varphi}$.

Tracing over the A-subsystem of the density matrix $\hat{\varrho}_{AB} = |F\rangle_{AB}\langle F|$, the arriving probabilities of the idler photon in $D_{Bx}$ and $D_{By}$ detectors become

$$P_{Bx} = {}_B^f\langle x|[Tr_A(\hat{\varrho}_{AB})]|x\rangle_B^f = \frac{1}{2}\left[1 - v\cos(\vartheta - \theta_g)\right], \tag{5}$$

$$P_{By} = {}_B^f\langle y|[Tr_A(\hat{\varrho}_{AB})]|y\rangle_B^f = \frac{1}{2}\left[1 + v\cos(\vartheta - \theta_g)\right], \tag{6}$$

where $v = \cos\left(\frac{\chi}{2}\right)$ is the interference visibility, and the phase $\theta_g = \frac{\chi}{2}$ is the geometric phase. The reason the phase shift $\theta_g$ is called geometric can be found in the ref. [20], and references therein. The authors of [20] proposed the operational approach for measuring the geometric phase in interferometry. In this approach, when a mixed or pure quantum state undergoing a cyclic or non-cyclic unitary evolution in one arm of the interferometer, the geometric phase can be measured as a shift in the dynamical oscillations of the output intensity. See also ref. [21] for more details.

Similarly, by tracing over the B-subsystem, the detection probabilities of the signal photon read

$$P_{Ax} = \frac{1}{2}\left[1 + v\cos(\varphi - \theta_g)\right], \quad P_{Ay} = \frac{1}{2}\left[1 - v\cos(\varphi - \theta_g)\right]. \tag{7}$$

The cross-Kerr nonlinearity can be measured from the geometric phase shift in the output interference pattern.

Before proceeding to the next section, it is important to note that in the standard quantum mechanics, the interference at the outputs of A- and B-MZIs can be controlled only "locally" through $\varphi$ and $\vartheta$ respectively. The interference effect of one subsystem is independent of the controlled (relative) phase applied on the other one, as it is expected.

## 3 Postselected interferometry: A special type of quantum erasers

In the previous example, each photon moving in the MZI represents a qubit. This allows to illustrate the peculiar effect of pre- and postselected ensembles using the simplest quantum mechanical system. The idler photon moving in B-MZI is the qubit system to be postselected, and A-MZI is the qubit-type measuring device which measures the phase changes with every successful postselection in B-MZI. For simplicity of the formulas, from now on, we shall assume that $\varphi = 0$, and $\chi \ll \pi$, except where indicated.

The postselection is performed by discarding all experimental runs at one port of B-MZI and taking into account the detection events at the other port, as in Figure 1. By postselecting the state $|y\rangle_B^f$, the density operator $\hat{\varrho}_{AB} = |F\rangle_{AB}\langle F|$, transforms into: $\hat{\varrho}_{AB} \longrightarrow \hat{\rho}_A \otimes \hat{\Pi}_{By}$, where $\hat{\Pi}_{By} = |y\rangle_B^f\langle y|$ is the projective operator. From the Formula 4, the output conditional density operator of the signal photon reads

$$\hat{\varrho}_A = |f\rangle_A\langle f|, \quad |f\rangle_A = \frac{1}{\sqrt{Z}}\left(\beta |x\rangle_A^f + \delta |y\rangle_A^f\right), \tag{8}$$

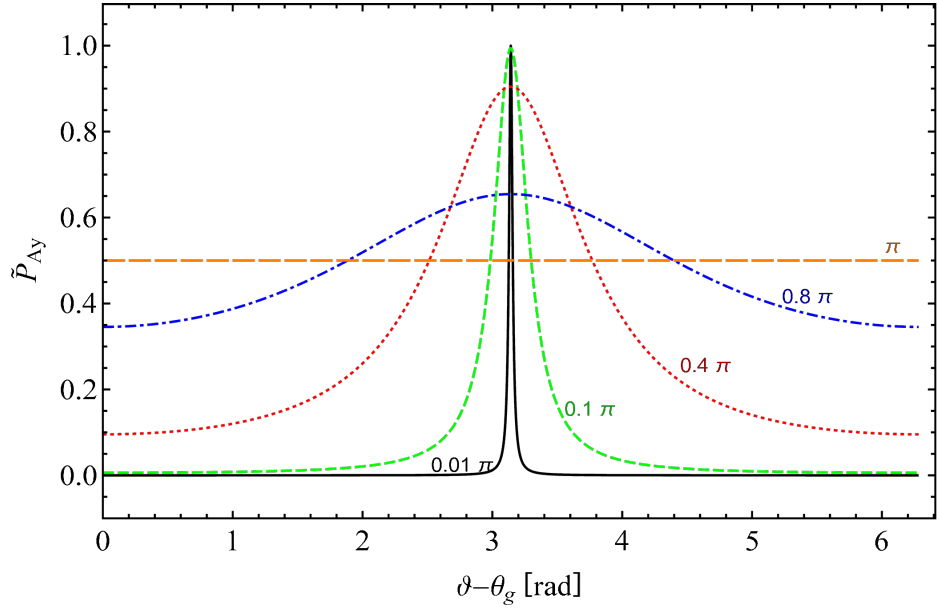

Figure 2: The output probability $\widetilde{P}_{Ay}$ at the y port of A-MZI versus the phase difference $\vartheta - \theta_g$ in B-MZI for different values of $\chi$. In the strong interaction regime, the visibility decreases and the interference completely disappear for $\chi = \pi$. In the weak interaction regime for $\chi \ll \pi$, the visibility $v = \cos\left(\frac{\chi}{2}\right) \simeq 1$ and the effect is localized in a small region around $\vartheta - \theta_g = \pi$, making the weak interaction regime perfect for quantum metrology.

where $Z =_A \langle f | f \rangle_A = \beta\beta^* + \delta\delta^*$ is the normalization factor. Assuming $\varphi = 0$, the output probabilities of A-MZI becomes

$$\widetilde{P}_{Ay} = \frac{\delta\delta^*}{Z} = \frac{\sin^2(\theta_g)}{2 + 2v\cos(\vartheta - \theta_g)}, \quad \widetilde{P}_{Ax} = \frac{\beta\beta^*}{Z} = 1 - \widetilde{P}_{Ay}. \tag{9}$$

By comparing the Expressions 9 and 7, one finds the postselected interferometry counter-intuitive. This follows from the ability of controlling the interference at the output of A-MZI by means of $\vartheta$ applied on B-MZI [22]. Figure 2 shows $\widetilde{P}_{Ay}$ as a function of the phase difference $(\vartheta - \theta_g)$ for different values of the cross-phase $\chi$. When $\vartheta - \theta_g = \pi$, the probe and reference states of B-MZI interfere out of phase along the postselected port y. At the same time, A-MZI measures a $\pi$ phase change.

This weird behavior of the quantum system under postselection may be classified as a special type of quantum erasers.

In most quantum eraser experiments [23], the postselection is done to exclude events that give information about the paths of the photon. In other words, the postselection erases the which-path information from the ensemble and recovers an interference pattern.

In the present model, the postselection also recovers an interference pattern. The experimenter postselects particular events at the output of B-MZI and observes the interference pattern ($\vartheta$-dependence) at the output of A-MZI.

However, the postselection used here is completely different from the one used in the quantum eraser. It results from the interference between the probe and reference states of B-MZI, and hence, it does not allow for any information about the paths of the postselected photon (idler photon). The purpose of applying it is to "erase" the effect of averaging over all possibilities at the output of B-MZI.[1]

---

[1]The erasing of the high-probability events allows studying the physical consequences of rare events; the behavior of the quantum system near the interference minima.

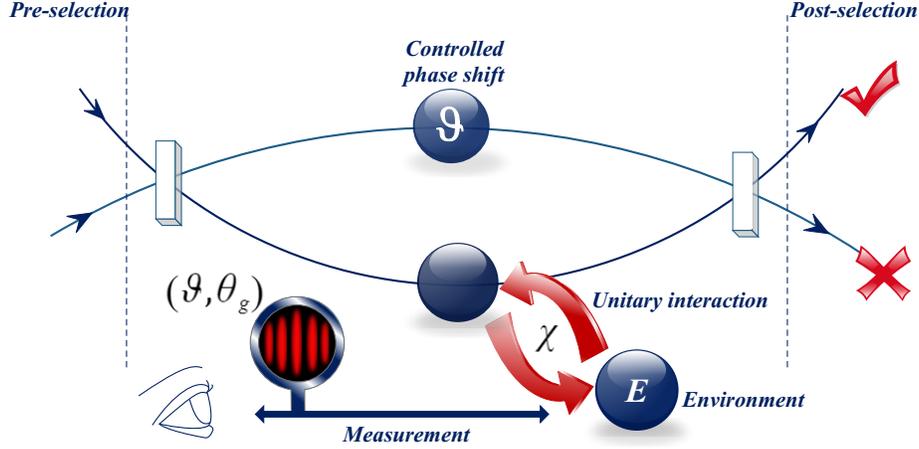

Figure 3: A special type of interference phenomenon characterizes the intermediate unitary interaction between the pre- and post-selected quantum system and the element of environment. The interference pattern can be controlled by a phase applied on the postselected system. It is a purely quantum mechanical effect and has no classical analogue.

This dependence of the outputs of A-MZI on $\vartheta$ applied in B-MZI is a completely hidden interference phenomenon in the standard treatment. Its recovering by postselection can be regarded as a quantum eraser-like effect, see Figure 3.

## 4  Optimal amplification of the effect of a single postselected photon

The amplification of the nonlinear effect of a single postselected photon has been studied using the weak-value amplification technique [3, 4].

In the theory in ref. [3], it has been also considered a system of two Mach-Zehnder interferometers coupled by cross-Kerr nonlinearity with some substantial changes from the model used in this paper. In one interferometer, single photons are injected and the rare events of postselections are prepared by using an imbalanced output beam splitter. In the other interferometer, a coherent state is used as an input. The authors showed that a single postselected photon can "act like many photons" in terms of the phase that it can impart to the coherent state interacting with it.[2]

This interpretation was derived from the calculation of the weak value of the photon number operator in the optical nonlinear medium: $\left\langle \hat{n}^r_{By} \right\rangle_w = \frac{1}{2\epsilon}$, where $\epsilon$ is a small real number related to the imbalanced reflectivity and transmissivity of the output beam splitter [3]. Note that the weak value of the photon number in the reference arm must gives: $\left\langle \hat{n}^r_{Bx} \right\rangle_w = 1 - \left\langle \hat{n}^r_{By} \right\rangle_w$. When $\epsilon \ll 1$, the weak value of the photon number in the probe arm of interferometer becomes $\left\langle \hat{n}^r_{By} \right\rangle_w \gg 1$. At the same time, in the reference arm of the interferometer, the weak value is $\left\langle \hat{n}^r_{Bx} \right\rangle_w \ll 0$, and the number of photons in this arm is negative!. However, the total number of photons in the interferometer is $\left\langle \hat{n}^r_{By} \right\rangle_w + \left\langle \hat{n}^r_{Bx} \right\rangle_w = 1$.

The experimental implementation in ref. [4] of the theory proposed in [3], used the same logic. The input state in the postselected system is a coherent state that contain one photon on average. The weak-value amplification technique gave an amplified phase of $\sim 50\mu rad$, which is eight times larger than the expected cross-phase in the standard case (i.e. without amplification). That is, a single postselected photon can act like eight photons [4].

---

[2]In the original work on the weak-value amplification in Ref. [2], a spin of a spin-$\frac{1}{2}$ particle can turn out to be 100.

We have seen in the previous section that the postselection induces a shift in the measuring device much larger than the magnitude of the single-photon nonlinearity $\chi$. However, this shift is not proportional to the weak value.

First, in the present model, the application of a phase shift with balanced beam splitters gives an imaginary weak value

$$\left\langle \hat{n}^r_{By} \right\rangle_w = \frac{{}^f_B\langle y|\hat{n}^r_{By}|x\rangle^i_B}{{}^f_B\langle y|x\rangle^i_B} = \frac{1}{1+e^{i\vartheta}}, \qquad (10)$$

instead of a real one, where $\hat{n}^r_{By}$ is the photon number operator in the probe arm of B-MZI. Therefore, a complex weak value is obtained and one cannot explain the shift of the measuring device using an anomalous real weak value.

Second, the weak value has no upper bound for the amplification. Both real (when $\epsilon \to 0$) and imaginary (for $\vartheta \to \pi$) weak values give anomalous and wrong predictions $\left\langle \hat{n}^r_{By} \right\rangle_w \to \infty$. Therefore, the interpretation of the shift of the measuring device using the weak value [3, 4, 14] is obviously nonphysical. According to the weak value, Formula 10, the amplification for $\vartheta = \pi$, is maximal. From Figure 2, the amplification is maximal only for out-of-phase interference when $\vartheta = \pi + \theta_g$, and for any other values of the controlled phase $\vartheta \neq \pi + \theta_g$, the amplification decreases.

To amplify the nonlinear effect of a single postselected photon to the value $\pi$, the input state must be a pair of time-correlated single photons. In place of the imbalanced output beam splitter, one has to apply a controlled phase $\vartheta$ with a balanced beam splitter. The controlled phase $\vartheta$ is then adjusted with great care to the value $\vartheta = \pi + \theta_g$. The inferred phase shift at the output of A-MZI can be estimated using the formula [24, 25]

$$\Theta(\vartheta - \theta_g) = \arccos\left(\frac{n^f_{Ax} - n^f_{Ay}}{n^f_{Ax} + n^f_{Ay}}\right) = \arccos\left(\frac{\beta\beta^* - \delta\delta^*}{\beta\beta^* + \delta\delta^*}\right), \qquad (11)$$

where, $n^f_{Ax} = n\widetilde{P}_{Ax}, n^f_{Ay} = n\widetilde{P}_{Ay}$ are the total number of photons detected in $D_{Ax}$ and $D_{Ay}$ for $n$ postselections. When $\varphi = \theta_g$, and $\vartheta = \pi + \theta_g$, the phase $\Theta = \arccos(-1) = \pi$. And a single "out-of-phase" postselected photon in B-MZI can impart a $\pi$ phase shift to the other photon moving in A-MZI regardless of the strength of the Kerr nonlinearity.

By comparing Figure 4 in the present work with Figure 4 in the ref. [4], one finds easily that the amplification effect of the weak-value amplification technique lies somewhere in the tail region of the function $\Theta(\vartheta - \theta_g)$.

To summarize this section, the shift in the measuring device (A-MZI) is induced by the phase difference $(\vartheta - \theta_g)$ between the interfered states at the output of the postselected system. The out-of-phase interference for $\vartheta - \theta_g = \pi$ represents the peak of the amplification effect that can be measured in A-MZI. What is interpreted as an amplification using the weak value is a pure phase effect and the weak value plays no role in its interpretation since it contains no information about the phase difference $(\vartheta - \theta_g)$.

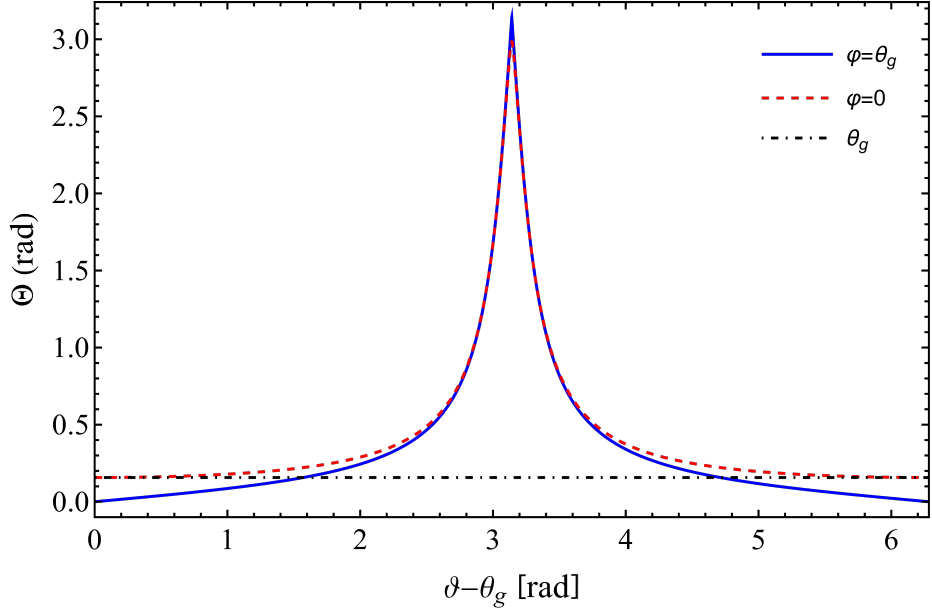

Figure 4: The phase-like variable $\Theta\left(\vartheta - \theta_g\right)$ in A-MZI with every postselection in B-MZI for $\chi = \frac{1}{10} rad$. The straight line represents the geometric phase shift $\theta_g$ of the interference pattern in the standard case. Under pre- and post-selection, the out-of-phase interference at the output of B-MZI for $\vartheta - \theta_g = \pi$, can impart an exact phase shift of $\pi$ to A-MZI irrespective of the strength of Kerr nonlinearity $\chi$. This amplification effect is standard (nonanomalous) and easy to control.

## 5 Parameter estimation

According to the weak-value amplification, the postselection is done on rare events [3, 4]. The postselection probability can be written as [14]

$$P = \left|\langle \psi^{post} | \psi^{pre} \rangle\right|^2 = \left|{}^f_B\langle y | x \rangle^i_B\right|^2 = \left|\frac{1 + e^{i\vartheta}}{2}\right|^2 = \frac{1}{2}(1 + \cos\vartheta) \qquad (12)$$

However, to perform an optimal parameter estimation, the postselection must be done on the "rarest" events.[3]

The postselection probability given by the Formula 12 is trivial. It cannot help in finding the rarest event of postselection since it describes the postselected quantum system (B-MZI) in the absence of the interaction, and hence, it contains no information about the geometric phase (unlike the Formulas 5 and 6).

From Formula 6, the realistic probability $P_{By}$ have a non-zero minimum, which corresponds to the "rarest" event we seek. For instance, by putting $\vartheta = \pi + \theta_g$, we get the lowest detection probability at the y port: $P_{By} = \sin^2\left(\frac{\chi}{4}\right) \neq 0$.

In the previous section, we have shown that the successful postselection with such low probability can lead to an optimal amplification. Anyway, it is not clear how the postselection on a rare event can be a useful tool in metrology. Next, we investigate how the entanglement is the only secret behind the effectiveness of any postselection-based metrological protocol.

---

[3]Intuitively, rare events always bring more information than high-probability ones. Shannon defined this fact by introducing a simple quantity called the self-information: $I = -\log P$, which measures the unexpectedness of an event [26], P is the probability.

## 5.1 Purification of the entanglement

Consider the general entangled state given by Formula 4. The aim is to purify this entangled state in a non-maximally entangled state of the form $|F\rangle \propto |x\rangle_A^f |x\rangle_B^f + \varepsilon |y\rangle_A^f |y\rangle_B^f$, where $|\varepsilon| \ll 1$. Due to the non-maximally entangled state, the two nearly dark ports of MZIs become completely correlated. The correlation between rare events on the output ports of MZIs can optimize the measurement signal-to-noise ratio.

The geometric phase plays a main role in the purification. According to the previous sections, the rarest event of out-of-phase postselection for $\vartheta = \pi + \theta_g$ in B-MZI can switch A-MZI by a $\pi$ phase, causing another rare event of the detection of the signal photon at the y port of A-MZI to deterministically occur ($\widetilde{P}_{Ay} = \frac{\delta\delta^*}{Z} = 1$) for $\varphi = \theta_g$. Substituting these values of $\varphi$ and $\vartheta$ into the Formula 4, the two-qubit output state for $\chi \neq \pi$ becomes

$$|F\rangle = -e^{i\frac{3\chi}{4}} \left[ \cos\left(\frac{\chi}{4}\right) |x\rangle_A^f |x\rangle_B^f - i \sin\left(\frac{\chi}{4}\right) |y\rangle_A^f |y\rangle_B^f \right]. \tag{13}$$

a non-maximally entangled state. In the limits, $\chi = \pi$, $and\ \chi = 0$, the previous state becomes

$$|F\rangle = \begin{cases} -\frac{e^{i\frac{3\chi}{4}}}{\sqrt{2}} \left[ |x\rangle_A^f |x\rangle_B^f - i |y\rangle_A^f |y\rangle_B^f \right], & \chi = \pi, \quad \text{maximally entangled state,} \\ |x\rangle_A^f |x\rangle_B^f, & \chi = 0, \quad \text{product state.} \end{cases} \tag{14}$$

The non-maximally entangled state is the resource for optimal parameter estimation in our model.

## 5.2 Postselected Fisher information

Fisher information is an index of how much information a distribution has about an unknown parameter. It is an important quantity to specify the uncertainty of measurements. Fisher information is defined as the negative second derivative of the log-likelihood function [27].

To determine the likelihood function, we consider $n$ postselections. Based on the click "1" and no-click "0" events, one can assign to every value of the controlled phase shift in the B-MZI $\vartheta_i : (\vartheta_1, \vartheta_2, \vartheta_3, \ldots)$ an observation $n_i^y : (n_1^y, n_2^y, n_3^y, \ldots) \in [0,n]$ represents the total number of clicks in $D_{Ay}$ detector in A-MZI. The number of clicks $N_i^y \leq n$ is a random variable obeys the binomial distribution $N_i^y \sim \text{Bin}(n, p_i)$, where $p_i(\vartheta_i - \theta_g)$ is the click probability, for simplicity we denote $\widetilde{P}_{Ay}$ by $p_i$. The likelihood law is thus given by:

$$L\left(n_i^y \mid p_i\right) = \binom{n}{n_i^y} (p_i)^{n_i^y} (1-p_i)^{n-n_i^y}. \tag{15}$$

Before deriving Fisher information. Maximum Likelihood (ML) Estimation method can be used to estimate $p_i$ which maximizes $L\left(n_i^y \mid p_i\right)$ [27]. Since the logarithm is a monotonic function, then maximizing $L\left(n_i^y \mid p_i\right)$ is equivalent to the maximizing of its log-likelihood $\log L\left(n_i^y \mid p_i\right)$. The log-likelihood reads:

$$l\left(n_i^y \mid p_i\right) = n_i^y \log(p_i) + \left(n - n_i^y\right) \log(1 - p_i) + const, \tag{16}$$

The ML estimator can be obtained from: $\frac{dl(p_i)}{dp_i}\bigg|_{p_i = \hat{p}_i^{ML}} = \frac{n_i^y}{p_i} - \frac{n - n_i^y}{1 - p_i} = 0$, which gives $\hat{p}_i^{ML} = \frac{n_i^y}{n}$ for a particular observation $N_i^y = n_i^y$. The expected value of $\hat{p}_i^{ML}$ is $\langle \hat{p}_i^{ML} \rangle = \frac{\langle N_i^y \rangle}{n} = \frac{1}{n} n p_i = p_i$, and its variance is $\sigma_i^2\left(\hat{p}_i^{ML}\right) = \frac{1}{n^2} . \sigma^2\left(N_i^y\right) = \frac{1}{n^2} n p_i (1 - p_i) = \frac{p_i(1-p_i)}{n}$.

Fisher information is given by the expected value of the negative second derivative of the function $l\left(N_i^y \mid p_i\right)$

$$F_n(p_i) = -\left\langle \frac{d^2 l\left(N_i^y \mid p_i\right)}{dp_i^2} \right\rangle = \frac{n}{p_i(1-p_i)}. \tag{17}$$

For a large number of postselections $n$, $p_i$ will approach the exact theoretical probability $\widetilde{P}_{Ay}$ given by Formula 9. When $\vartheta_i = \pi + \theta_g$, Fisher information becomes

$$F_n(\theta_g) = \frac{n}{\widetilde{P}_{Ay}\left(1-\widetilde{P}_{Ay}\right)} = \frac{n}{\sin^2\left(\frac{\chi}{4}\right)\cos^2\left(\frac{\chi}{4}\right)} = \frac{4n}{\sin^2(\theta_g)}, \tag{18}$$

This is the maximum accessible Fisher information because of the perfect correlation with the $x$ output port of A-MZI.

## 5.3 Conservation of information under the postselection

In the standard quantum mechanics, all experimental runs (quanta) are taken into account. Under the pre- and post-selection procedure and weak interaction $\chi$, only a tiny portion of runs is considered.

When $\vartheta = \pi + \theta_g$, the mean number of runs to get a single out-of-phase postselected photon is $R = \frac{1}{P_{By}} = \frac{1}{\sin^2\left(\frac{\chi}{4}\right)} \simeq \frac{16}{\chi^2}$. Therefore, for n postselections, the total number of discarded runs in B-MZI is $nR - n \simeq nR$. At the same time, the expected Fisher information in A-MZI, given by Equation 18, can be increased by the same amount $F_n \simeq nR$.

Clearly, discarding runs by postselection may cause no loss of information. The out-of-phase postselection can increase Fisher information by the same amount as the sample's reduction caused by the postselection in B-MZI. And all information can be compressed in the out-of-phase postselected events $n \ll nR$.

## 5.4 Uncertainty and signal-to-noise ratio

The maximal Fisher information $F_n = nR$ means that the uncertainty in measuring the geometric phase shift can not be found below the standard quantum limit. If $\tau$ is the data acquisition time and $\Gamma$ is the photon pairs input rate, the total number of postselections becomes $n = P_{By}\Gamma\tau$, and the maximal Fisher information

$$F_n = \frac{16 P_{By} \Gamma \tau}{\chi^2} \simeq \Gamma\tau, \tag{19}$$

depends on the single-photon input rate in the postselected system. In interferometry, the quantity of interest is the unknown phase. In the present work, this phase quantity is the geometric phase shift, see Equation 9. From the Cramér-Rao bound [28, 29], the maximal phase sensitivity

$$\Delta\theta_g \geq \frac{1}{\sqrt{F_n}} = \frac{\chi}{4}\frac{1}{\sqrt{n}} = \frac{1}{\sqrt{\Gamma\tau}}, \tag{20}$$

is at the standard quantum limit. The signal-to-noise ratio (SNR) is used to quantify this phase sensitivity. SNR is defined as the expectation value of the measured quantity to its standard deviation. For $n$

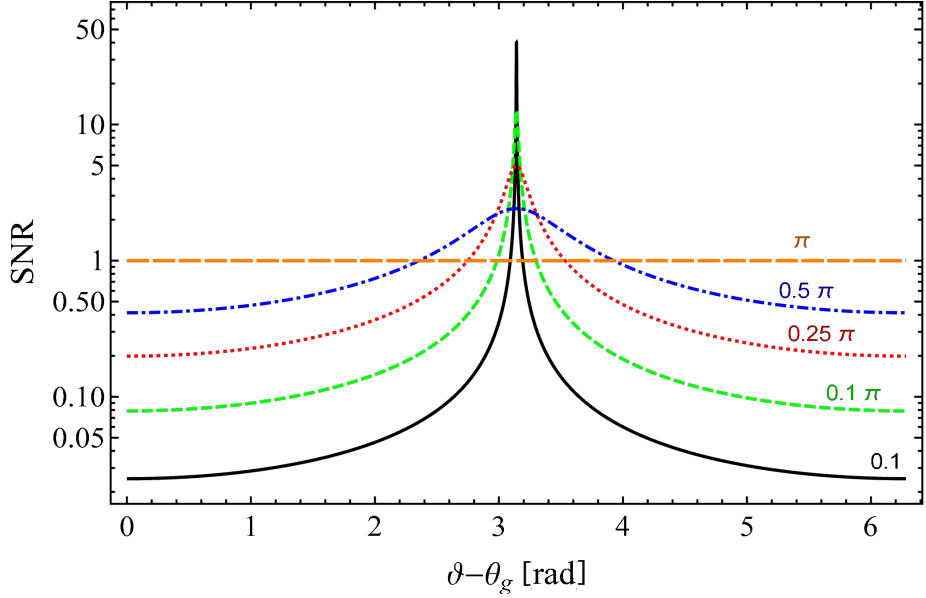

Figure 5: SNR as a function of the phase difference $\vartheta - \theta_g$ for different values of interaction strength of the cross-Kerr nonlinearity $\chi$ and single postselection $n = 1$. SNR can be improved by out-of-phase interference for $\vartheta - \theta_g = \pi$ and weak interaction $\chi$. For small displacements of $\vartheta$ from $\theta_g$, SNR rabidly degrades.

postselections the SNR gives [30]

$$SNR = \frac{\langle N_i^y \rangle}{\sigma\left(N_i^y\right)} = \frac{n\widetilde{P}_{Ay}}{\sqrt{n\widetilde{P}_{Ay}\left(1 - \widetilde{P}_{Ay}\right)}} \equiv \sqrt{n}\sqrt{\frac{\widetilde{P}_{Ay}}{\widetilde{P}_{Ax}}}, \tag{21}$$

where $n$ is the total number of postselections (detected signals) and $\sqrt{\widetilde{P}_{Ay}/\widetilde{P}_{Ax}} = \sqrt{\delta\delta^*/\beta\beta^*}$ is the optimization factor. When $\vartheta = \pi + \theta_g$, $SNR = \sqrt{n} \cot\left(\frac{\chi}{4}\right) \simeq \sqrt{\Gamma\tau}$, and SNR is improved in such a way that no matter how low the detected signals are, the noise vanishes and the SNR is determined by the square root of total input photons $\sqrt{\Gamma\tau} \gg \sqrt{n}$.

Practically, such enhancement in SNR expresses a great deal of confidence that almost no detection event in the nearly dark port $D_{Ay}$, upon every low-probability postselection in $D_{By}$, can result from a coincidence with any noise factor (dark or readout noises) in $D_{Ay}$, but it is a signal comes from the pure correlation between these rare events (i.e. between the nearly dark ports of MZIs).

See Figure 5 and compare this improvement in the SNR near $\theta_g$ with weak-value amplification techniques in Refs. [3, 31].

## 6 Some final notes

In this work, we have used the so-called instantaneous model of the cross-Kerr non-linearity [19], assuming an ideal transformation through Kerr interaction. This model is only valid for the cases where the nonlinearity has a response time that is much longer than the single-photon pulse duration [32]. In the process of parametric down-conversion, signal and idler photons birth simultaneously but leave the crystal as wave packets with temporal widths of the order of $15 fs$ [33]. Optical fibers have a response time of about $50 fs$ [34]. Therefore, the instantaneous model can be applied for such optical materials. In general, the condition for the proposed scheme to be feasible in experiments is that the temporal width of the single-photon wave packets must be shorter than the response time of the optical medium to keep the noise in the induced phase low.

The no-go theorem in ref. [32] states the impossibility of getting a $\pi$ phase shift in the fast response regime due to the noisy nature of Kerr interaction for single-photon wave packets. The present results may offer a method to avoid this no-go theorem. The imparted $\pi$ phase by the out-of-phase postselected photon to the other photon interacting with it comes from a quantum effect but not from the strength of interaction between photons, which can take any value.

For quantum metrology applications, the present model offers a method to measure the geometric phase. From the geometric phase, one of the unknown phases can be determined precisely, either the controlled phase shift or the nonlinear cross-phase $\chi$.

In the standard interferometry, all quanta must be detected. The detectors are saturated for any input rate of photons bigger than $10^7$ counts per second [35]. This imposes a limitation on the input power.

In postselected interferometry, we deal with rare events without losing any of the information that can be acquired from ideal standard interferometry (see section 5.3). Most input quanta that leave the bright port of the interferometer are not detected, and the detector's saturation is no more a limitation.

Not detecting most of the input quanta on the output does not mean the discarding of them completely. Due to the non-maximally entangled state 13, the photons leaving bright $x$ ports of the MZIs can be completely recycled [36] by a feedback mechanism (reinjecting them to the MZIs).

For photons with wavelengths $\lambda = 1\mu m$, and photon pairs input rate $\Gamma \simeq 10^{11} s^{-1}$, the input power is about $\sim 10 nW$. If the data acquisition time is $\tau = 10\ hours$, the uncertainty in measuring the geometric phase from the Relation 20 gives $\Delta\theta_g \simeq 10^{-8} rad$.

If the cross-phase shift $\chi$ is known precisely [37], the uncertainty in measuring longitudinal phase shift can reach $\Delta\vartheta = \Delta\theta_g = 10^{-8} rad$ for the aforementioned input rate and data acquisition time. This corresponds to a sensitivity in the displacement of the phase shifter $\vartheta$ of about $x = \frac{\Delta\vartheta \lambda}{2\pi} \simeq 1 fm$, which is very close to the approximated proton radius [38].

To reduce the noise coming from the dark or readout noise in $D_{By}$ detector, a superconducting single-photon detector with milli-Hz dark counts rate can be used [39].

Another factor that can help in improving the measurement is the using of a fast-response nonlinear material. Such materials allow using high input rate of photon pairs. Here, optical fibers are the best candidates. They are promising tools for a wide range of quantum applications due to their unique properties of ultra-flattened dispersion, ultra-low loss, and ultra-fast response time [40, 41].

Comparing with the input power used in the weak value amplification [12, 42], the present model allows achieving high-sensitive measurements using low input power. Therefore, enhancing the sensitivity of postselected metrology requires not only the weak interaction but also the operation at the point $\pi + \theta_g$, which corresponds to the rarest event of postselection.

# 7 Conclusion

In summary, optimal amplification and estimation of small effects under pre- and post-selection requires i) the preselection of single-particle Fock states, ii) the intermediate weak interaction between a qubit-type quantum system and a qubit-type meter, and iii) the postselection of events characterized by out-of-phase interference (depending on the geometric phase shift of the interference pattern).

It is instructive to compare the results of the proposed model with the weak-value amplification technique. The comparison may include three main issues: the interpretation of the quantum effects of

pre- and post-selected ensembles, the amplification of the nonlinear effect of a single postselection, and the estimation of small parameters.

In interpretation issues. Instead of using the controversial quantity of weak value, the peculiar quantum effects of pre- and post-selected ensembles can be interpreted as a quantum eraser-like effect. The postselection of the quantum states erases the effect of averaging over different alternatives, and consequently, recovers a completely hidden interference phenomenon in the measuring device.

In the amplification. We have shown by studying the interference pattern that a single photon resulting from "out-of-phase" interference in the postselected system can impart a $\pi$ phase shift to the other photon interacting with it. This is the peak of amplification. It is several orders of magnitude more than the weak-value amplification [4, 3].

In the estimation of small parameters. The arranging for out-of-phase postselection events is the optimal strategy for quantum postselected metrology. A pair of photons, initially prepared in a product state, can be entangled in a non-maximally entangled state through an intermediate weak interaction and a subsequent merging of out-of-phase quantum states. The non-maximally entangled state is a strong correlation between rare events that can help in perfectly surpassing the "technical" noise.

Finally, the studying of pre- and post-selected ensembles as a quantum eraser-like effect can provide novel insights into the quantum world, offer a more comprehensive understanding of it, and show how to improve our current quantum sensors.

# Achnowledgment

I thank Prof. Leonid Il'ichov and Dr. Vladimir Tomilin for valuable discussions and for reading the draft of this paper and providing helpful feedback. This work was performed at the Institute of Automation and Electrometry of the Siberian Branch of the Russian Academy of Sciences within the framework of the State Assignment (Project No. AAAA-A17-117052210003-4, internal FASO number 0319-2016-0002).